\documentclass{mem}
\usepackage{natbib}\usepackage{txfonts}\usepackage{balance}
\usepackage{graphicx}
\bibpunct{(}{)}{;}{a}{}{,} 
\idline{75}{282}
\begin{document}

\title{
THESEUS in the era of Multi-Messenger Astronomy
}

   \subtitle{}

\author{
G.\, Stratta\inst{1,2,3} 
\and L.\, Amati\inst{3}
\and R.\,Ciolfi\inst{4,5}
\and S.\, Vinciguerra\inst{6}
          }

\institute{
Universit\'a degli studio di Urbino "Carlo Bo" -- 
Dipartimento di Scienze di Base e Fondamenti, Via S. Chiara 27, 
61029, Urbino, Italy
\email{giulia.stratta@uniurb.it}
\and
INFN, Sezione di Firenze, via G. Sansone 1, I-50019 Sesto Fiorentino, Firenze, Italy
\and
INAF --
Osservatorio Astronomico e Scienze dello Spazio, Via Gobetti,
I-34131 Bologna, Italy
\email{lorenzo.amati@iasfbo.inaf.it}
\and
INAF -- 
Osservatorio Astronomico di Padova, Vicolo dell'Osservatorio 5, 
I-35122 Padova, Italy
\and
INFN-TIFPA -- Trento Institute for Fundamental Physics and Applications, via Sommarive 14, I-38123 Trento, Italy
\and
Institute of Gravitational Wave Astronomy \& School of Physics and
Astronomy, University of Birmingham, Birmingham, B15 2TT, United Kingdom
}

\authorrunning{Stratta }

\titlerunning{THESEUS and the  Multi-Messenger Astronomy}

\abstract{
The recent discovery of the electromagnetic counterpart of the gravitational wave source GW170817  
has demonstrated the huge informative power of multi-messenger observations. 
Late '20s and early '30s will be a mature era for multi-messenger astronomy. 
Consolidated  network of second generation gravitational wave detectors, 
such as Advanced LIGO and Advanced Virgo, KAGRA and LIGO-India, will be further powered by the contribution from third generation interferometers such as Einstein Telescope and/or Cosmic Explorer. Several astrophysical sources detectable in GWs are expected to radiate in the full electromagentic spectrum and to emit high energy neutrinos, thus requiring a robust synergy with ground- and space-based high energy detectors (e.g. CTA, THESEUS, ATHENA), sensitive neutrino detectors (e.g. KM3Net,  IceCube-Gen2) and large size optical facilities (e.g. E-ELT). In this report we review the fundamental role of THESEUS in this exciting context. 
}

\maketitle{}

\section{Introduction}

Stellar-mass black hole coalescences, 
together with binary neutron star (NS-NS), NS-black hole (BH) mergers, and burst sources as core-collapsing massive stars and 
possibly NS instability episodes are among the main targets of ground-based gravitational waves (GWs) detectors, an ensemble of Michelson-type 
interferometers sensitive to the high frequency range, from few Hz to few thousand Hz. 
Some of these transient sources are also expected to produce neutrinos and electromagnetic (EM) signals over the entire spectrum, 
from radio to gamma-rays.  
Since 2015, the Universe in GWs has been discolosed with the 
first detection of GWs from black hole binary systems during their  coalescing phase 
\citep{Abbott2016b, Abbott2016a}. 
On August 17$^\mathrm{th}$, 2017, the detection of a GW signal 
consistent with a binary neutron star merger system \citep{Abbott2017a} and shortly preceding the short gamma-ray 
burst GRB170817A \citep{Abbott2017b} marked the first simultaneous detection of 
GW and EM radiation from the same source.  At the same time, 
significant evidence of high-energy (TeV-PeV) cosmic neutrinos has recently been obtained from an extensive IceCube fourth-year 
data analysis \citep[e.g.][]{Aartsen2014b}. The origin of these neutrinos is still unknown but the lack of significant anisotropy 
in the data sky direction distribution is consistent with an (at least partially) extragalactic origin of the neutrino sources. 

\begin{figure*}[t!]
\centering
\includegraphics[scale=0.45,angle=90]{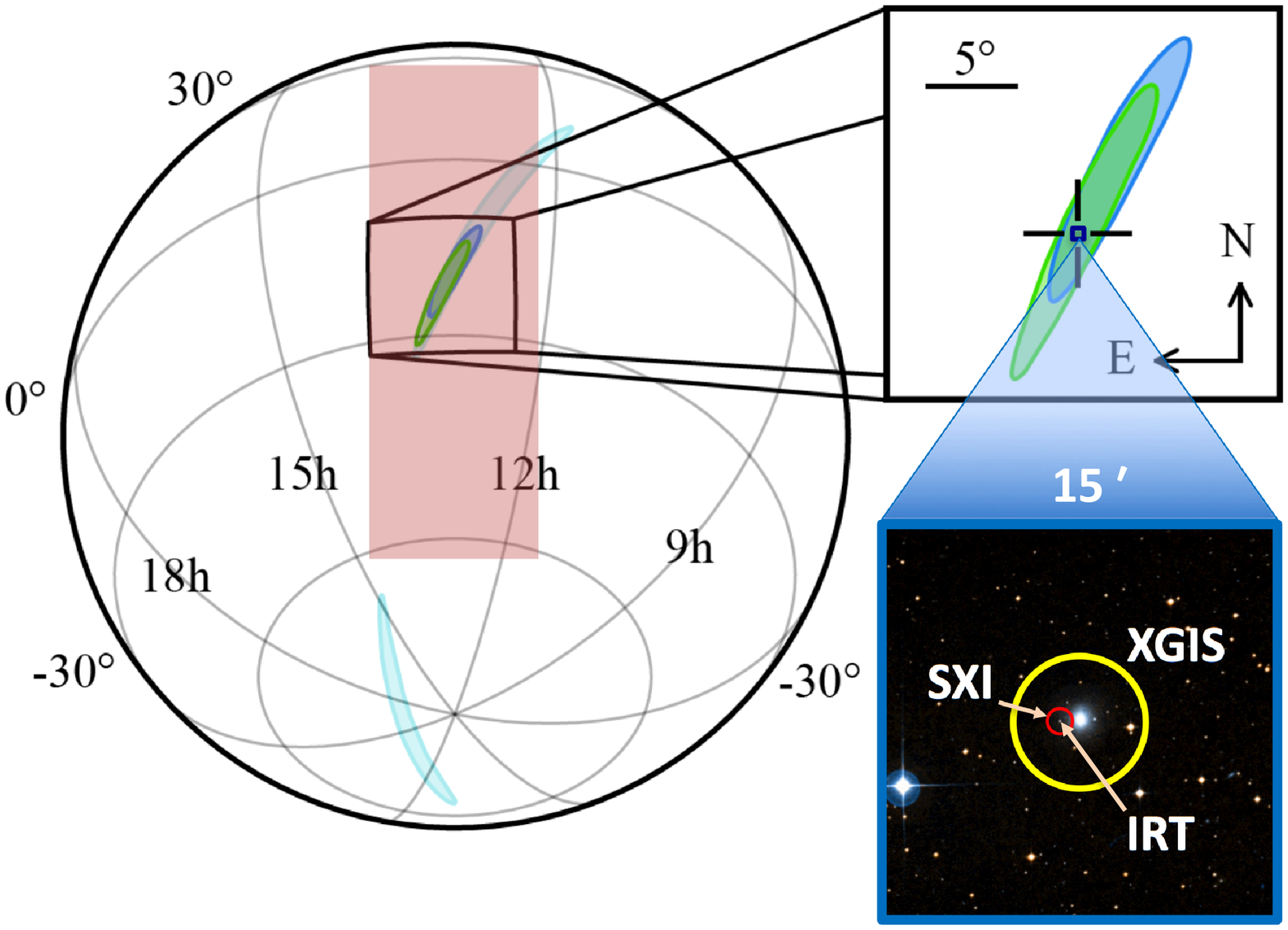}
\caption{
The plot shows the THESEUS/SXI field of view ($\sim 110\times30$~deg$^2$, pink rectangle) superimposed on the probability skymap of 
GW 170817 obtained with the two Advanced LIGO only (cyan) and with the addition of Advanced Virgo (green) \citep{Abbott2017a}. 
THESEUS not only will cover a large fraction of the skymap (even those obtained with only two GW-detectors, e.g. cyan area), but will 
also localize the counterpart with uncertainty of the order of 5 arcmin with the XGIS and to less than 1 arcmin with SXI. 
{\it The THESEUS location accuracy of GW events produced by NS-NS mergers can be as good as 1~arcsec in case of detection 
of the kilonova emission by the IRT}. By the end of the 2020s, if ET will be a single detector, almost no directional 
information will be
available for GW sources ($>1000$ deg$^2$ for BNS at $z>0.3$, \citealt[e.g.][]{Zhao2017}), and a GRB-localising 
satellite will be essential to discover EM counterparts.}
\label{fig:SXI_FOV_GW170817}   
\end{figure*}

The detection of electromagnetic (EM) counterparts of GW and neutrino signals enables a multitude of science 
programmes by allowing for parameter constraints that the GW or neutrino observations alone cannot fully provide 
\citep[see, e.g.,][]{Bloom2009, Phinney2009}. 
Each individual joint observation of an EM source and its GW and/or neutrino counterpart, provides an enormous science return.  
For example, in the case of compact binary coalescences, the determination of the GW polarization ratio would 
constrain the binary orbit inclination and hence, when combined with an EM signal, the jet geometry and source energetics.  
A better understanding of the NS equation of state can follow from the combined detection of GWs and X-ray signals  
\citep[see, e.g.,][]{Bauswein2012,Takami2014,Lasky2014,Ciolfi2015a,Ciolfi2015b,Messenger2015,Rezzolla2016,Drago2016}. 
Multi-messenger observations will likely answer to the questions whether the magnetar scenario and 
the current interpretation of X-ray giant flares are correct.  
The identification and characterization of the kilonova emission associated with NS-NS or NS-BH coalescences will 
enable to trace the history of heavy-metal enrichment of the Universe.  
Cosmological redshift measurements of a large sample of short GRBs combined with the absolute source luminosity distance 
provided by the CBC-GW signals can deliver precise measurements of the Hubble constant \citep{Schutz1986}, 
helping to break the degeneracies in determining other cosmological parameters via CMB, SNIa and BAO surveys \citep[see, e.g.,][]{Dalal2006}. 

\begin{figure*}
\centering
\includegraphics[scale=0.50]{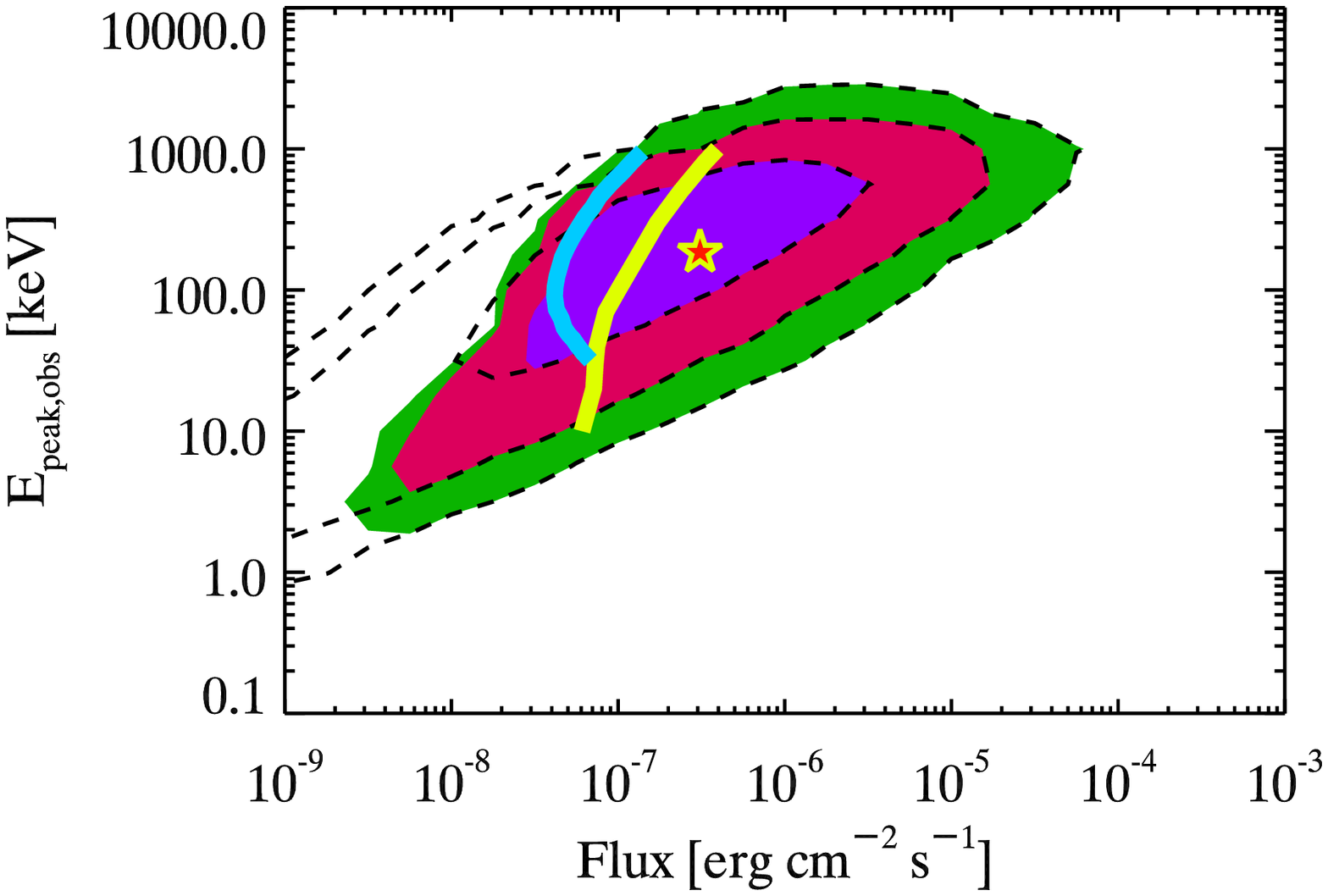}
\vspace{3cm}
\caption{Density contours (dashed lines) corresponding to  1, 2, 3 $\sigma$ levels of the synthetic population of Short GRBs 
(from \citealt{Ghirlanda2016}). Shaded coloured regions show the density contours of the population detectable by THESEUS. 
The yellow and cyan lines show the trigger threshold of Fermi/GBM and GCRO/Batse (from \citealt{Nava2011}). 
The flux is integrated over the 10-1000 keV energy range. The star symbol shows the short GRB170817A \citep{Goldstein2017}.}
\label{fig:capabilities}
\end{figure*} 

By the end of the twenties, the sky will be routinely monitored by the second-generation GW detector network, 
composed by the two Advanced LIGO (aLIGO) detectors in the US \citep{Harry2010}, Advanced Virgo (aVirgo) 
in Italy \citep{Acernese2015}, ILIGO in India \citep[e.g.][]{Abbott2016} and KAGRA in Japan \citep{Somiya2012}. 
More sensitive third generation GW detectors, such as the Einstein Telescope \citep[ET,  e.g.][]{Punturo2010} 
and LIGO Cosmic Explorer \citep[LIGO-CE, e.g.][]{Abbott2017d}, are planned to be operational in the 2030s and to provide an increase of 
roughly one order of magnitude in sensitivity. In parallel to these advancements, IceCube and KM3nNeT and the advent of 
10 km$^3$ detectors \citep[e.g. IceCube-Gen2,][and references therein]{Aartsen2014} will enable to gain high-statistics 
samples of astrophysical neutrinos.  Therefore, the '30s will be a golden era of Multi-Messenger Astronomy (MMA). 

GW detectors have relatively poor sky localisation capabilities, mainly based on triangulation methods, that on average will not be better than 
few dozens of square degrees \citep{Abbott2016}. 
For GW sources at distances larger than the horizon of second-generation detectors (200 Mpc), therefore accessible only by the third-generation 
ones in the 2030s  (e.g. Einstein Telescope, ET,  and Cosmic Explorer, LIGO-CE \citealt{Punturo2010,Abbott2017b}), sky localization may even worsen 
if the new generation network will be composed by only one or two detectors, with possible values of the order of few hundred 
square degrees or more \citep[e.g.][]{Zhao2017}. 
Neutrino detectors can localise to an accuracy of better than a few squares degrees 
\citep[see, e.g.,][and references therein]{Santander2016}. 
In order to maximise the science return of the multi-messenger investigation of these transient sources it is essential to have 
a facility that (i) can detect and disseminate an EM signal independently to the GW/neutrino event and (ii) can rapidly search 
with good sensitivity in the large error boxes provided by the GW and neutrino facilities.  These combined requirements 
are uniquely fulfilled by THESEUS ({\it Transient High Energy Sky and Early Universe Surveyor}\footnote{http://www.isdc.unige.ch/theseus}). 

\section{The role of THESEUS in the MMA}
THESEUS is a mission concept developed by a large International collaboration currently under evaluation by ESA within the selection process 
for next M5 mission of the Cosmic Vision Programme \citep{Amati2017}. 
In the context of multi-messenger observations, THESEUS will trigger and localize transient sources within the uncertain GW and/or neutrino error boxes with the X-Gamma ray Imaging Spectromater (XGIS, 2 keV - 20 MeV)  and/or with the Soft X-ray Imager (SXI, 0.3 - 6 keV). A very large fraction of the error boxes of poorly localised GW sources can be covered with SXI field of view (FoV, see Fig.\ref{fig:SXI_FOV_GW170817}) within one orbit due to the large grasp of the instrument (see \citealt{Amati2017}).  In response to an SXI/XGIS 
trigger, if an optical counterpart is present, the source sky localization can be refined down to few arcseconds 
with onboard fast repointement of the infrared telescope (IRT). Precise localizations will be disseminated within minutes to the astronomical community, 
thus enabling large ground-based telescopes to observe and deeply characterise the transient nature. 

\begin{table*}
\scriptsize
\label{tab:rate}  
\begin{tabular}{llllll}
\hline\noalign{\smallskip}
\multicolumn{3}{c}{GW observations} &  \multicolumn{3}{c}{THESEUS XGIS/SXI joint GW+EM observations} \\
\noalign{\smallskip}\hline\noalign{\smallskip}
Epoch & GW detector & BNS range & BNS rate & XGIS/sGRB rate & SXI/X-ray isotropic \\
& & & (yr$^{-1}$) & (yr$^{-1}$) & counterpart rate  (yr$^{-1}$) \\
\noalign{\smallskip}\hline\noalign{\smallskip}
2020+ & Second-generation (advanced LIGO, & $\sim$200~Mpc & $\sim$40$^*$ & $\sim$5-15 & $\sim$1-3 (simultaneous) \\ 
& Advanced Virgo, India-LIGO, KAGRA) & & & & $\sim$6-12 (+follow-up) \\
2030+ & Second + Third-generation & $\sim$15-20~Gpc & $>$10000 & $\sim$15-35 & $\gtrsim$100 \\
& (e.g. ET, Cosmic Explorer) & & & &  \\
\noalign{\smallskip}\hline
\end{tabular}
\note*{from Abadie et al. 2010a}
\end{table*}

Several multi-messenger sources are among the 
main targets of THESEUS, as for example GRBs, flaring magnetars, core-collapse supernovae (CCSNe) and AGNs \citep[e.g.][]{Stratta2017,Amati2017}. 
In particular, the combination of SXI and XGIS, makes THESEUS a unique machine to explore both the populations of long/high redshift and hard/short GRBs. 
Figure \ref{fig:capabilities} shows the density contours of the population of short GRBs (dashed contours) in the peak energy - peak flux 
plane \citep{Ghirlanda2016}. The density contours of the short GRB population detectable by THESEUS is shown by the shaded contours. 
Due to their harder spectrum, short GRBs are better triggered by XGIS than SXI. 
Compared to the detection thresholds of BATSE and Fermi/GBM, THESEUS will slightly extend the detected population leftwards of these 
thresholds (cyan and yellow lines in Figure \ref{fig:capabilities}). The plot also shows the position 
of GRB170817A as revealed by GBM \citep{Goldstein2017}. THESEUS will be able to fully access similar events and explore their nature.  
Although the XGIS sensitivity threshold improves over GBM, its smaller (by a factor of 2) field of view compensates this gain reaching 
a detection rate of short GRBs which is comparable to that of GBM.  What makes THESEUS XGIS unique, with respect to GBM, is the possibility 
to locate, thanks to the soft (2 keV-30 keV) coded mask detectors of the XGIS, most of the detected short GRBs with an expected accuracy 
of 5 arcmin (to be compared with the average $>$ few degrees of GBM GRBs).

\begin{figure*}[t!]
\centering
\includegraphics[scale=0.125]{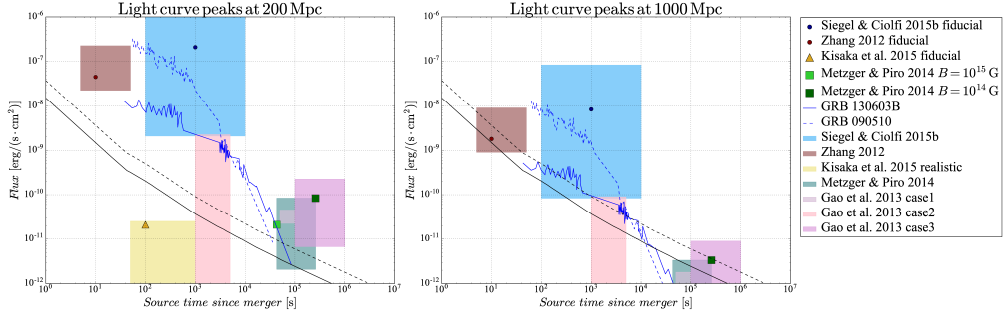}
\caption{Expected X-ray fluxes at peak luminosity, with corresponding uncertainties, from different models of magnetar-powered X-ray emission from long-lived NS-NS merger remnants, computed assuming two different luminosity distances (z=0.05 on the left panel, and z=0.2 on the right panel). 
Predictions from each model are represented by a coloured region and/or by single dots that are indicative of fiducial cases (see the legend on the 
right). Grey solid lines in the left panel show typical GRB X-ray afterglows observed with Swift/XRT. The black curves show the SXI 
sensitivity vs. exposure time, assuming a source column density of $5\times10^{20}$ cm$^{-2}$ (i.e., well out of the Galactic plane 
and very little intrinsic absorption, solid line) and $10^{22}$ cm$^{-2}$ (significant intrinsic absorption, dashed line). 
}
\label{fig:X-rayFlux}   
\end{figure*}

THESEUS will ensure short GRB detection with a rate of 15-35 per year \citep{Stratta2017}. 
The short GRB detection capabilities highlight the crucial relevance of the role of THESEUS for multi-messenger astronomy in an epoch where almost all short GRBs will be accompanied by a GW signal detected by the third-generation interferometers (e.g. ET or LIGO-CE).  
Besides the expected collimated GRB ``prompt" emission, softer X-ray emission is also 
expected from the afterglow component as observed so far in more than 90$\%$ of GRBs.  The afterglow of a XGIS detected GRB 
can be followed-up with SXI. 

Table \ref{tab:rate} shows the expected rate of THESEUS/XGIS short GRB detections with a GW counterpart from merging NS-NS systems (i.e. within the GW detector horizon). The quoted numbers are obtained by correcting the realistic estimate of NS-NS GW detection rate, $\sim40$ yr$^{-1}$ 
(\citealt{Abadie2010a}, see also \citealt{Belczynski2017}) for the fraction of the sky covered by the XGIS FoV, that is $\sim$50\%, and 
the short GRB jet collimation factor, by assuming a jet half-opening angle range of 10--40 deg, taking into account the possibility to 
observe off-axis short GRBs up to 5 times a jet half-opening angle of 10 degrees\footnote{The 5-times factor was obtained by considering 
the THESEUS/XGIS 1 sec photon flux sensitivity $\sim$0.2 ph cm$^{-2}$ s$^{-1}$ (see also \citealt{Amati2017})} and 2 times a jet of 40 
degrees \citep{Kathirgamaraju2017,Pescalli2016}. We are here assuming that every BNS merger produces a jetted short GRB, which is still an open issue. 
Results show that, during the 2020's, the GW+EM detection rate of short GRBs  with THESEUS is found to be of the order of 5--15 per year. 

By the time of the launch of THESEUS, gravitational radiation from NS-NS and NS-BH mergers will be detectable by third-generation detectors 
such as the Einstein Telescope (ET) up to redshifts z$\sim$2 or larger \citep[see, e.g.,][]{Sathyaprakash2012, Punturo2010}, thus dramatically 
increasing the GW+EM on-axis short GRB detection rate. 
The important implication is that almost all THESEUS short GRBs will have a detectable GW emission. Indeed, it is likely that at the typical 
distances at which ET detects GW events, the only EM counterparts that could feasibly be detected are short GRBs and their afterglows, 
making the role of THESEUS crucial for multi-messenger astronomy by that time. 

GW emission from CBCs depends only weakly on the inclination angle of the inspiral orbit and therefore these events are in general observable 
at any viewing angle. As a consequence, most of the GW-detected mergers are expected to be observed off-axis (i.e. with a large angular 
distance of the observer from the orbital axis). This makes the non collimated, nearly isotropic EM components extremely relevant for 
the multi-messenger investigation of CBCs. 
A potentially powerful nearly-isotropic emission is expected if a NS-NS merger produces a long-lived millisecond magnetar as well as off-axis 
afterglow emission. Figure \ref{fig:X-rayFlux} shows the expected X-ray fluxes at peak luminosity from different models of magnetar-powered X-ray emission from long-lived NS-NS merger remnants. 
The expected detection rate of the isotropic X-ray emission from NS-NS mergers is quoted in the last column of Table \ref{tab:rate} where, from the realistic rate 
of NS-NS mergers that will be detected with GW observatories in 2020s and 2030s, we have accounted for: 1) the fraction of the sky covered by 
the SXI FoV, that is $\sim8$\%, for serendipitous discoveries, and 2) the fraction of NS-NS systems that can produce X-ray emission (i.e. 
that do not form immediately a BH), that we assumed to be within [30-60]\% \citep{Gao2013,Piro2017}. Moreover, we consider the fraction 
of BNS sources that could be followed-up with SXI after a GW alert, estimated to be of the order of $\sim40$\%.  
From these computations, we find that during the 2020s the joint GW+EM detection rate with THESEUS of these X-ray counterparts of 
NS-NS mergers is $\sim$ 6-12 per year. During 2030s, with the third-generation GW detectors, isotropic X-ray emission from NS-NS mergers 
as predicted by some models \citep[e.g.][]{Siegel2016b} could be detected up to $\sim10$ times larger distances, with an improved 
joint GW+EM detection rate of few hundreds per year (depending on the largely uncertain intrinsic luminosity of such X-ray component, 
see Figure \ref{fig:X-rayFlux}). With such statistics, THESEUS will provide a unique contribution to characterize 
this X-ray emission from NS-NS systems.

Both serendipitous discoveries within the large THESEUS/SXI FoV and re-pointing of THESEUS in response to a GW trigger will 
allow to study X-ray emission expected from NS-NS systems. With THESEUS/SXI in 
combination with the second-generation detector network, almost all predicted non-collimated X-ray counterparts of GW events 
from NS-NS merging systems will be easily detected simultaneously with the GW trigger and/or with rapid follow-up of the 
GW-individuated sky region. Among the open questions that THESEUS will help to address there are: 1) does the NS-NS merger 
create a NS or a BH, and how fast?; 2) how much matter is expelled in the NS mergers? At which speeds?; 3) what is the amount 
of asymmetry in the NS-NS merger ejecta and the corresponding optical emission?
\begin{figure*}[t]
\centering
\includegraphics[scale=0.50]{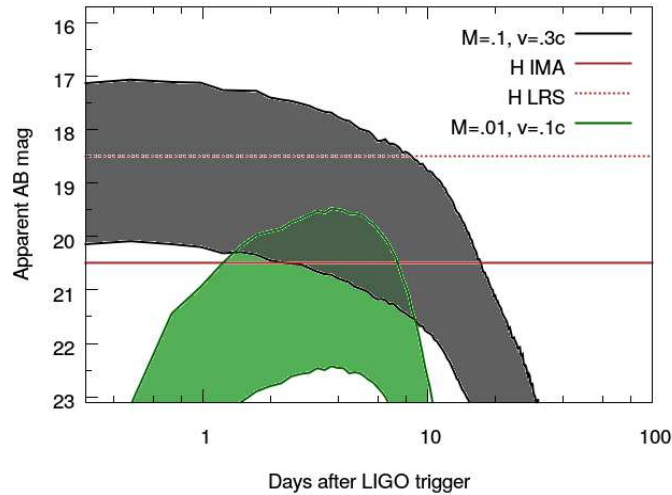}
\caption{
Theoretical $H$-band lightcurves of kilonova based on models \citep[from][]{Barnes2016}. The lightcurves are in observer 
frame for a source between 50 and 200 Mpc. Gray model is for the most optimistic case of a kilonova with $0.1\, M_\odot$ 
ejected mass with speed of $0.3\,c$. Green model is for a weaker emission, corresponding to $0.01\, M_\odot$ ejected mass 
with speed of $0.1\,c$. The continuous and dashed red lines indicate the THESEUS/IRT limiting H magnitudes for imaging 
and prism spectroscopy, respectively, with 300 s of exposure \citep[see][]{Amati2017}. }
\label{fig:kilonova}   
\end{figure*}

Another well known type of nearly-isotropic emission expected from CBCs involving NSs is the so-called ``kilonova'' or 
``macronova'' \citep[e.g.,][]{Li_Paczynski1998,Metzger2010}. 
Figure~\ref{fig:kilonova} shows how the infrared telescope onboard THESEUS (IRT), due to its promptly follow-up capabilities,   
would enable a detailed characterization of these key astrophysical phenomena associated with compact 
binary coalescences and short GRBs.  Specifically, the plot shows the expected macronova apparent magnitudes 
for a source between 50 and 200~Mpc \citep{Barnes2016} plotted against the IRT sensitivity in imaging and 
spectroscopic mode. If bright enough, spectroscopic observations could be performed on-board, 
thus providing redshift estimates and information on chemical composition of circum-burst medium. 
Alternatively, if the optical/NIR counterpart is not bright enough for onboard spectroscopy, 
precise IRT sky coordinates will be disseminated to ground based telescopes to perform spectroscopic 
observations and redshift measurements.

\section{Conclusions}

The first detection of the electromagnetic counterparts of a GW source has confirmed a number of theoretical expectations and 
boosted the nascent multi-messenger astronomy. Several classes of sources, including compact 
binary coalescences discussed in this review, as well as core-collapsing massive stars, and instability episodes on NSs that are expected to originate simultaneously 
high-frequency GWs, neutrinos and EM emission across the entire EM spectrum, including in particular high energy emission 
(in X-rays and gamma-rays). The mission concept THESEUS has the potential to play a crucial role in the 
multimessenger investigation of these sources. 
THESEUS, if approved, will have the capability to detect a very large number of transient sources in the X-ray and gamma-ray 
sky due to its wide field of view, and to automatically follow-up any high energy detection in the near infrared. 
In addition, it will be able to localize the sources down to arcminute (in gamma and X-rays) or to arcsecond (in NIR). 
As we have shown in this paper,  the instrumental characteristics of THESEUS are ideal to operate in synergy with 
the facilities that will be available by the time of the mission: several new generation ground- and space-based 
telescopes, second- and third-generation GW detector networks and 10 km$^3$ neutrino detectors. This makes THESEUS 
perfectly suited for the coming golden era of multi-messenger astronomy and astrophysics.

\begin{acknowledgements}
S.V.: the research leading to these results has received funding from the
 People Programme (Marie Curie Actions) of the European Union's Seventh
Framework Programme FP7/2007-2013/ (PEOPLE-2013-ITN) under REA grant
agreement no.~[606176]. This paper reflects only the authors' view and the
European Union is not liable for any use that may be made of the
information contained therein.
\end{acknowledgements}

\bibliographystyle{aa}
\bibliography{theseus_mma} 

\end{document}